\documentclass[english,aps,prl,reprint,longbibliography,notitlepage,superscriptaddress,preprintnumbers,twocolumn,showkeys]{revtex4-2}

\usepackage{graphicx}
\usepackage{bm}
\usepackage{hyperref}
\usepackage{amsmath}
\usepackage{amssymb}
\usepackage{microtype}
\usepackage{xcolor}
\usepackage[capitalise]{cleveref}
\usepackage{bbm}
\usepackage{mathtools}
\usepackage{dutchcal}
\usepackage{tensor}
\usepackage[normalem]{ulem} 

\renewcommand*{\vec}[1]{\bm{#1}}
\newcommand*{\unitvec}[1]{\vec{\hat{#1}}}
\newcommand*{\sci}[1]{\ensuremath{\times 10^{#1}}}
\newcommand*{\pfactor}[1]{\textcolor{blue}{\ensuremath{\times{#1}}}}

\DeclareMathAlphabet{\mathpzc}{OT1}{pzc}{m}{it}

\newcommand{\para}[1]{\par\vspace{2mm}\noindent\textbf{#1}\,---\,}

\makeatletter
\DeclareRobustCommand{\rcite}[1]{%
  \rcite@aux#1,\@nil{#1}%
}
\def\rcite@aux#1,#2\@nil#3{%
  \if\relax#2\relax
    Ref.~\cite{#3}%
  \else
    Refs.~\cite{#3}%
  \fi
}
\makeatother

\hypersetup{
    colorlinks = true,
    citecolor = {red},
    linkcolor = {blue},
    urlcolor = {blue},
}

\newcommand{\LCDM}{$\Lambda$CDM}

\newcommand{\WMAP}{WMAP}
\newcommand{\Planck}{{\textit{Planck}}}
\newcommand{\Nside}{\ensuremath{N_{\mathrm{side}}}}

\begin{document}

\title{Strong Evidence Against a Statistically Isotropic Universe}

\author{Joann Jones}
\email{joannjones@uchicago.edu}
\affiliation{Kavli Institute for Cosmological Physics, University of Chicago, Chicago, IL 60637, USA}
\affiliation{Department of Astronomy \& Astrophysics, University of Chicago, Chicago, IL 60637, USA}

\author{Craig J. Copi}
\email{craig.copi@case.edu}
\affiliation{CERCA/ISO, Department of Physics, Case Western Reserve University, 10900 Euclid Avenue, Cleveland, OH 44106, USA}

\author{Glenn D. Starkman}
\email{glenn.starkman@case.edu}
\affiliation{CERCA/ISO, Department of Physics, Case Western Reserve University, 10900 Euclid Avenue, Cleveland, OH 44106, USA}

\author{Yashar Akrami}
\email{yashar.akrami@csic.es}
\affiliation{Instituto de F\'isica Te\'orica (IFT) UAM-CSIC, C/ Nicol\'as Cabrera 13-15, Campus de Cantoblanco UAM, 28049 Madrid, Spain}
\affiliation{CERCA/ISO, Department of Physics, Case Western Reserve University, 10900 Euclid Avenue, Cleveland, OH 44106, USA}

\date{\today}

\begin{abstract}
The standard cosmological model predicts statistically isotropic cosmic microwave background (CMB) fluctuations characterized by the CMB temperature coefficients $a_{\ell m}$ being independent Gaussian random variables with zero mean and with variance that depends only on $\ell$.
However, several summary statistics of CMB isotropy   have anomalous values, including: the low level of large-angle temperature correlations, $S_{1/2}$;
the excess power in odd versus even low-$\ell$ multipoles, $R^{TT}$; the (low) variance of large-scale temperature anisotropies in the ecliptic north, but not the south, $\sigma^2_{16}$; and the alignment and planarity of the quadrupole and octopole of temperature, $S_{QO}$.
Individually, their low $p$-values are weak  evidence for violation of statistical isotropy.   
We study the tail values of these statistics and find very little correlation among them. We show that the joint probability of all four anomalies occurring by chance in \LCDM\ is likely $\leq3\times10^{-8}$. We examine the balance in the impact of look-elsewhere effects and the existence of other anomalies on the significance of this result. We argue that non-Gaussianity alone is  unlikely to account for the anomalies seen at the level of the angular power spectrum, $C_\ell$, and that they instead appear to require correlations between $a_{\ell m}$.
Our results provide strong evidence for a violation of statistical isotropy, and we conclude that the anomalies should not be dismissed as flukes within \LCDM.
\end{abstract}

\preprint{IFT-UAM/CSIC-23-132}

\maketitle

\para{Introduction.}
Observations of the cosmic microwave background (CMB) temperature and polarization anisotropies can rightfully be regarded as the strongest evidence in support of a nearly spatially flat, statistically isotropic $\Lambda$ Cold Dark Matter (\LCDM) model of the Universe.
However, there are several large-angle features of the CMB, known as the large-angle anomalies, that suggest violations of statistical isotropy \cite{Schwarz:2015cma,Planck:2013lks,Planck:2015igc,Planck:2019evm,Abdalla:2022yfr}.
Each feature is characterized by an anomaly-specific statistic $S_a$ having a $p$-value, or the probability of making an observation at least as extreme as that actually observed, low enough to excite notice. Typically $p(S_a)<0.01$ in an appropriate ensemble of \LCDM~realizations.
Each anomaly was first identified in either the Cosmic Background Explorer (COBE) \cite{Hinshaw:1996ut} or Wilkinson Microwave Anisotropy Probe (WMAP) data \cite{WMAP:2003elm}, and has subsequently been confirmed in more recent \Planck\ data (starting from \rcite{Planck:2013lks}).  
Within \LCDM\ these anomalies are separate from the parameter ``tensions'' that have received much attention, such as the $H_0$ tension \cite{Abdalla:2022yfr}. 

Though each anomaly has a low $p$-value, they are often excused as statistical flukes due to their identification \textit{a posteriori} and look-elsewhere effects. This is true if all the anomalies are correlated with one another within \LCDM. However, if they are uncorrelated, their joint $p$-value may actually be far smaller than their individual ones, and far harder to excuse. 
In this Letter, we consider four anomalies: the absence of large-angle correlations, characterized by $S_{1/2}$; the parity asymmetry ($R^{TT}$); the low northern variance ($\sigma^2_{16}$, as a way of characterizing hemispherical asymmetry); and the quadrupole-octopole alignment ($S_{QO}$).
We explore whether or not these anomalies are correlated, i.e., how their joint probability distribution compares to the products of their individual probability distributions.

While the correlations between anomaly statistics have previously been studied \cite{Muir:2018hjv}, since each observation has a low $p$-value, we are not interested in how the peaks of the distributions correlate -- instead, we must study the tails of the distributions, which are not captured by standard correlation measures.
Here we show  that the joint $p$-values of all pairs of these four anomalies in an ensemble of realizations of best-fit \LCDM~are much less than the minima of the individual $p$-values,
\begin{equation}
    p(S_A, S_B) \ll \min[p(S_A), p(S_B)]\,, 
\end{equation}
for any pair of anomaly statistics $S_A$ and $S_B$. The statistics are only weakly correlated in their tails.

We find that for three of the four \Planck\ foreground-subtracted maps -- \texttt{Commander}, \texttt{NILC}, \texttt{SMICA} --
the joint $p$-value of all four anomalies 
\begin{equation}
    p(S_{1/2},R^{TT},\sigma^2_{16}, S_{QO}) \leq 3\sci{-8} \,.
\end{equation}
The outlier \texttt{SEVEM} map still has $p = 1.8\sci{-7}$. In those three maps, the $p$-value of the large-angle anomalies is less than that of a $5.4\sigma$ Gaussian fluctuation. 
There are also other anomalies that one expects would reduce this $p$-value still further.
Although there are undoubtedly look-elsewhere penalties to be paid for this collection of mostly \textit{a posteori} statistical anomalies,
such strong evidence that we do not live in a realization of a Gaussian random statistically isotropic $\Lambda$CDM universe clearly demands further study.

It is conventional to express both the predicted and observed CMB temperature  on the sky in terms of an expansion in spherical harmonics:
\begin{equation}
    \label{eqn:Tthetaphitoalm}
    T(\theta,\phi) = \sum_{\ell=0}^\infty\sum_{m=-\ell}^{\ell} a_{\ell m} Y_{\ell m}(\theta,\phi)\,.
\end{equation}
The $\ell=0$ term represents the average temperature, 
while $\ell\geq1$ are the temperature anisotropies, $\Delta T(\theta,\phi)$.
The $\ell=1$ term is dominated by a large ``Doppler'' contribution due to our motion relative to some appropriate average frame of rest, thus we conventionally focus our attention on the cosmologically relevant terms $\ell\geq2$.

\LCDM\ as conventionally constructed predicts that the $a_{\ell m}$ are statistically independent Gaussian (or nearly so) random variables of zero mean that reflect the  rotation and translation invariance of both the underlying theory and the background geometry, so that their expected correlations are purely diagonal (they are uncorrelated) and their variance depends only on $\ell$,
\begin{equation}
    \label{eqn:SIexpectations}
    \langle a_{\ell m} a_{\ell'm'}^* \rangle = C_{\ell}\delta_{\ell \ell'}\delta_{m m'} .
\end{equation}
Here $\langle\cdots\rangle$ denotes an average over an ensemble of realizations of the underlying  theory.
The anomalies we discuss in this paper call into question the assumptions and expectations encoded in \cref{eqn:SIexpectations}.

The $C_{\ell}$ in \cref{eqn:SIexpectations} are the theoretical angular power spectrum. 
In \LCDM, they encode (nearly) all relevant cosmological information.
They depend on the 7 cosmological parameters that characterize the homogeneous background geometry and its evolution, the stress-energy content of the Universe, and the phenomenological power spectrum of primordial fluctuations in the stress-energy and geometry.
Similar descriptions apply to polarization, but this work is confined to temperature fluctuations.
 
Enormous effort has been invested in building and operating CMB experiments so as (primarily) to extract estimates of the underlying $C_{\ell}$.
Simplistically, this is done by obtaining a full-sky map $T(\theta,\phi)$, extracting the $a_{\ell m}$, and computing the angular power spectrum 
\begin{equation}
    \label{eqn:Clfromalm}
    C_{\ell} \equiv \frac{1}{2\ell +1} \sum^{\ell}_{m = -\ell}  |a_{\ell m}|^2 \,.
\end{equation}
In practice, there are many subtleties \cite{Planck:2019nip,Planck:2018yye} including the presence of foregrounds and the consequent need to mask parts of the sky where foreground removal is unreliable.

The real-space angular correlation function is an alternative to the angular power spectrum:
\begin{equation}
    \label{eqn:ACF}
    C(\theta) = \langle T(\unitvec{n}_1)T(\unitvec{n}_2)\rangle 
        = \sum_{\ell=2}\frac{2\ell+1}{4\pi} C_{\ell}P_{\ell}(\cos\theta)\,,
\end{equation}
where $\unitvec{n}_1$ and $\unitvec{n}_2$ are two directions on the sky, with $\cos\theta=\unitvec{n}_1\cdot\unitvec{n}_2$, and $P_{\ell}$ are the Legendre polynomials. Strictly, the sum should start from $\ell=0$, but for the reasons given above we consider the monopole-and-dipole-subtracted angular correlation function.
In a statistically isotropic universe with $a_{\ell m}$ that are independent Gaussian random variables of zero mean, on a full sky the $C_{\ell}$ in \cref{eqn:ACF} are the same as the $C_{\ell}$ in \cref{eqn:Clfromalm}. 

\para{Maps and masks.}
We make use of the data maps, masks, and best-fit theoretical $\Lambda$CDM power spectrum provided by the \Planck\ 2018 data release found on the \Planck\ Legacy Archive \footnote{Available from \url{https://pla.esac.esa.int/#maps}.}. We use the component-separated \texttt{Commander}, \texttt{NILC}, \texttt{SEVEM}, and \texttt{SMICA} maps and apply the `common mask' to all maps. 
Since the calculation of one of the anomalies discussed below, $\sigma^2_{16}$, is based on the northern ecliptic sky, all maps and the mask are rotated to ecliptic coordinates before calculating any statistic. 
Since we are interested in large-scale features, we work with maps downgraded to \texttt{HEALPix} \footnote{See \url{http://healpix.sourceforge.net}.} resolution $\Nside = 64$ or to $\Nside = 16$. 
The rotating and downgrading process is as follows for the \Planck\ maps provided at $\Nside = 2048$. 
We first rotate each map to ecliptic coordinates using the \texttt{healpy} routine \texttt{healpy.rotator.Rotator}. 
We then downgrade the map to $\Nside = 64$ (for $S_{1/2}$ and $R^{TT}$) or to $\Nside=16$ (for $\sigma^2_{16}$) including smoothing with the appropriate Gaussian beams and applying the corresponding pixel window functions as described in \rcite{Muir:2018hjv}, i.e., we employ Gaussian beams with FWHM of $160$ arcminutes and $640$ arcminutes for $\Nside=64$ and $\Nside=16$, respectively.
We follow the same procedure for the masks, and choose a cutoff value of $0.9$ for resetting pixel values to 0 or 1.

\para{Creating realizations.}
We create $10^8$ \LCDM\ realizations of the CMB temperature using the \Planck\ best-fit cosmological parameters \cite{Planck:2018vyg} \footnote{Available from \url{https://pla.esac.esa.int/#cosmology}.}
up to $\ell_{\mathrm{max}}=200$.
To remain consistent with the data maps, we generate our simulated maps in ecliptic coordinates at $\Nside = 64$ or $\Nside=16$, as required, including the pixel window function and Gaussian beam smoothing appropriate for each $\Nside$.
Here we work with CMB-only simulated sky maps, i.e., we do not add simulated noise nor foreground to our realizations of the CMB sky. 
(See below for detailed discussions.)

\para{The anomalies.}
Broadly, the CMB anomalies are statistics other than the $C_\ell$ that have values far into the tails of their distribution in the ensemble of \LCDM\ realizations. 
Many statistics have been studied and many have been found to be anomalous, i.e., have a small $p$-value.
Typically the anomalous statistics that get reported have $p$-values less than a few percent, though none have individual $p$-values smaller than approximately $10^{-4}$. 

Each of these anomalies provides evidence against the validity of \cref{eqn:SIexpectations}, 
i.e., against statistical isotropy and, or, the Gaussianity of the Universe.
However, they are all \textit{a posteori} -- invented after looking at the data set in which they were first found to be anomalous.
This situation has led to something of a standoff --
there are now many examples of anomalies that cast statistical isotropy into doubt;
however, there is not yet a comprehensive fundamental model that post-dicts these anomalies and makes other testable predictions. 
(Cosmic topology may be an exception \cite{COMPACT:2022gbl}.)
Further, some have argued that physical models for anomalies are likely to make certain generic testable predictions for CMB polarization~\cite{Dvorkin:2007jp,Yoho:2015bla,Mukherjee:2015wra} and other observables \cite{Yoho:2013tta,Copi:2013cya}.

Little attention has been paid to the collective statistical strength of the anomalies.
The authors of \rcite{Muir:2018hjv} studied the correlations of certain CMB anomaly statistics, but their attention focused on the peaks of those distributions. 
In the following, we first recall a selection of four representative CMB anomalies and then determine the correlation in the tails of their \LCDM\ distributions.

We selected one representative statistic for each of four apparently unrelated phenomena, namely the lack of large-angle correlations, preference for odd parity, difference between northern and southern hemispheres, and alignment of the quadrupole and octopole. We emphasize that the
four statistics we have chosen are not exhaustive of the list of all anomalous statistics.
We closely follow the choices of \rcite{Muir:2018hjv} in order to ameliorate any suspicion that we have cherry-picked our results. There is no look-elsewhere penalty to be paid for selecting, unless perhaps we had been careful to choose the most anomalous member of each class. We refer the reader to \rcite{Muir:2018hjv} for in-depth discussions of many observational and statistical issues associated with these statistics and the maps on which they are calculated. Other statistics might certainly be added to this suite, such as the ones discussed in \rcite{Copi_2004,Land:2005ad,Yoho:2010pb,Chiocchetta:2020ylv,Pranav:2023khq}.

\paragraph{Lack of large-angle correlations, $S_{1/2}$:}
The lack of large-angle correlations in the CMB temperature anisotropies was first observed by COBE \cite{1993PNAS...90.4766B}, quantified by WMAP \cite{WMAP:2003elm,Copi:2006tu,Copi:2008hw}, and most precisely measured and characterized by \Planck\ \cite{Planck:2013lks,Copi:2013cya,Planck:2015igc,Schwarz:2015cma,Planck:2019evm}.
This deficit is quantified by the $S_{1/2}$ statistic: the squared distance of the two point angular correlation function $C(\theta)$ from zero for points on the sky with angular separations from $60^{\circ}$ to $180^{\circ}$,
\begin{equation}
    \label{eqn:S12}
    S_{1/2} \equiv \int_{-1}^{1/2}[C(\theta)]^2 \, \mathrm{d}(\cos\theta) \,. 
\end{equation}

Though $S_{1/2}$ can be calculated on full or masked skies, we follow the choices made by previous anomaly studies and only consider the masked sky statistic in this work.
 $S_{1/2}$ is calculated based on pseudo-$C_\ell$ extracted from $\Nside=64$ maps in ecliptic coordinates with the common mask applied using \texttt{NaMaster} \footnote{Code available from \url{https://github.com/LSSTDESC/NaMaster}.} \cite{Alonso:2018jzx} up to $\ell_{\mathrm{max}}=40$.
See \rcite{Copi:2008hw} for a discussion of its computation. 

Since $C(\theta)$, and hence $S_{1/2}$, is a function only of $C_\ell$ it is not automatically a test of statistical isotropy. 
However, we can rewrite $S_{1/2}$ as a quadratic form in the $C_\ell$ \cite{Copi:2008hw}:
\begin{equation}
    S_{1/2} = \frac{1}{4\pi^2}\sum_{\ell,\ell'=2}^\infty (2\ell+1)(2\ell'+1) C_{\ell} {\mathcal{I}}_{\ell\ell'} C_{\ell'} \,,
\end{equation}
where ${\mathcal{I}}_{\ell\ell'}$ is a calculable matrix \cite{Copi:2008hw}.
One way to achieve a low value of $S_{1/2}$ would have been for the first few low-$\ell$ $C_\ell$ to be unexpectedly small. 
This could have been explained, for example, by a suppressed low-$k$ power spectrum as a result of a modification of the inflaton potential. 
However, this is not why $S_{1/2}$ is small in our Universe -- only $C_2$ is significantly smaller than its expected value. 
Rather, as discussed in detail in \rcite{Copi:2008hw}, the contributions of $C_\ell$ with $\ell=2,3,4,5$   to $S_{1/2}$ cancel the contributions due to the higher $\ell$. 
(We refer here exclusively to masked-sky statistics.)
It is very difficult to arrange for this cancellation if $C_\ell$ are the variances of $2\ell+1$ independent zero-mean $a_{\ell m}$, even if the $a_{\ell m}$ are not Gaussian distributed.
Either the $a_{\ell m}$ need to have non-zero means of approximately the observed $\sqrt{C_\ell}$ and small variances  or the $a_{\ell m}$ of a given $\ell$ have to be correlated to drive down the variance of $C_{\ell}$, or the $a_{\ell m}$ have to be correlated across $\ell$s, in order to correlate the $C_\ell$. Each of these is a sign of statistical isotropy violation.

\paragraph{Parity asymmetry, $R^{TT}$:}
The parity asymmetry can be characterized in numerous ways.
Here we choose $R^{TT}$, a statistic characterizing the odd-parity preference at low $\ell$ in the CMB \cite{Kim_2010},
\begin{equation}
    R^{TT} \equiv \frac{D_+(\ell_{\mathrm{max}})}{D_-(\ell_{\mathrm{max}})}\,,
\end{equation}
where, choosing $\ell_{\mathrm{max}}$ to be odd,
\begin{equation}
    D_{\pm} \equiv \frac{2}{\ell_{\mathrm{max}}-1}\sum_{\ell = 2}^{\ell_{\mathrm{max}}}\frac{\ell(\ell +1)}{2\pi}
        \frac{(1\pm(-1)^\ell)}{2}C_{\ell}\,.
\end{equation}
Following \rcite{Muir:2018hjv}, we choose $\ell_{\mathrm{max}}=27$ (which is not the $\ell$ that minimizes the $p$-value) and calculate $R^{TT}$ on unmasked $\Nside=64$ skies. 

Again, since $R^{TT}$ is a function only of $C_\ell$ its anomalously low value is not automatically a measure of the violation of statistical isotropy, specifically the parity transformation of $O(3)$. 
However, it is again very difficult to see how this can be done in the way that the observed sky does -- by having $C_{2j+1}>C_{2j}$ consistently for $j=1,2,...,9$ -- while having the $C_\ell$ be the variances of (not-necessarily Gaussian) independent, random $a_{\ell m}$ of zero mean.

\paragraph{Low northern variance, $\sigma_{16}^2$:}
There are several indications that there is a statistically significant difference between the northern and southern hemispheres of the CMB sky. 
This was possibly first noted as an asymmetry in power \cite{Eriksen:2003db} that was maximized when the plane separating the hemispheres was approximately the ecliptic, and independently using genus statistics in \rcite{Park:2003qd}.
The anomaly was also characterized as ``dipole modulation'' of the power \cite{Gordon:2005ai} and as ``variance asymmetry''~\cite{Akrami:2014eta}.
\Planck\ 2013 \cite{Planck:2013lks} showed, using both full-resolution  ($\Nside = 2048$) and low-resolution ($\Nside = 16$) maps that the low variance was localized in the northern ecliptic hemisphere, with a $p$-value of $\sim0.1\%$, while the $p$-value for the southern hemisphere was $\sim45\%$.
We adopt the $\Nside = 16$ northern variance measure
\begin{equation}
     \sigma^2_{16} \equiv  \overline{(T - \bar{T})^2},
\end{equation}
where the overbar denotes an average over the pixels of an $\Nside = 16$ masked, northern ecliptic sky.

\paragraph{Quadrupole-octopole alignment, $S_{QO}$:}
There have been at least three ways presented to describe the alignment of the quadrupole and octopole \cite{deOliveira-Costa:2003utu,Schwarz:2004gk,Land:2005ad}.  
Following \rcite{Muir:2018hjv}, we adopt the approach of \rcite{Schwarz:2004gk}, which uses the multipole vector representation of CMB temperature \cite{Maxwell:1865zz,Copi:2003kt}. 
The quadrupole is represented by two multipole vectors $\unitvec{v}^{(2; j)}$, with $j=1,2$, and the octopole by three multipole vectors $\unitvec{v}^{(3; j)}$, with $j=1,2,3$.  
This allows the quadrupole to be associated with a single plane and the octopole to be associated with three planes.
These planes have oriented areas
\begin{equation}
     \vec{w}^{(\ell; i, j)} \equiv \unitvec{v}^{(\ell; i)} \times \unitvec{v}^{(\ell; j)}\,.
\end{equation}

It has been known since \rcite{deOliveira-Costa:2003utu} that the octopole is unusually planar -- its three planes are unexpectedly aligned with one another -- and  those planes are unusually aligned with that of the quadrupole.
We employ the $S_{QO}$ statistic to quantify the alignment of these four planes,
\begin{equation}
    S_{QO} \equiv \frac{1}{3} \sum_{i=1}^{2} \sum_{j=i+1}^{3} | \vec{w}^{(2; 1,2)}\cdot \vec{w}^{(3; i,j)}|\,.
\end{equation}

\para{Results.}
The value of each of $S_{1/2}$, $R^{TT}$, $\sigma_{16}^2$, and $S_{QO}$ for the four \Planck\ maps is given in the second column of \cref{tab:pairwisep}.  
Along the diagonal of columns 3--6 and for rows 1--4 of each component-separated map we present the $p$-values of those statistics from an ensemble of $10^8$ theoretical \Planck-best-fit \LCDM\ CMB skies. This is computed by determining the number of realizations with a given statistic at least as anomalous in the same direction as the data in order to capture the behavior of the tail of the distribution.
Each of those $p$-values is $\lesssim5\%$. 
In the off-diagonal entries below the diagonal we present the $p$-values of pairs of statistics;
in the off-diagonal entries above the diagonal (blue) we present the factors by which these exceed the products of individual $p$-values.
For example, in \texttt{SMICA} $p(S_{1/2},R^{TT})=2.8\times10^{-5}$, while  $p(S_{1/2},R^{TT})/(p(S_{1/2})p(R^{TT}))=0.6$. Note that consistently, only $p(S_{1/2},\sigma_{16}^2)/(p(S_{1/2})p(\sigma_{16}^2))\simeq30\gg1$; all other pairs have factors which are relatively close to $1$.  
$R^{TT}$, in particular, is very weakly correlated, or even anti-correlated, with the other three.
This is our first indication that the joint probability of all four statistics, is much less than the smallest of them.

\begin{table}[ht]
\begin{ruledtabular}
\begin{tabular}{llcccc}
Stat.& Value & $S_{1/2}$ & $R^{TT}$ & $\sigma_{16}^2$ & $S_{QO}$\\ \hline
\multicolumn{6}{c}{\texttt{Commander}}\\
\hline
$S_{1/2}$&1272 & $1.5\sci{-3}$ & \pfactor{0.6} & \pfactor{27} & \pfactor{1.3} \\
$R^{TT}$&0.7896 & $2.8\sci{-5}$ & $3.0\sci{-2}$ & \pfactor{1.1} & \pfactor{1.0} \\
$\sigma_{16}^2$&617.6 & $1.2\sci{-4}$ & $1.0\sci{-4}$ & $3.1\sci{-3}$ & \pfactor{1.7} \\
$S_{QO}$&0.7630& $8.3\sci{-6}$ & $1.3\sci{-4}$ & $2.3\sci{-5}$ & $4.4\sci{-3}$\\
\hline
\multicolumn{6}{c}{\texttt{NILC}}\\
\hline
$S_{1/2}$&1218 & $1.3\sci{-3}$ & \pfactor{0.4}  & \pfactor{29} & \pfactor{1.3} \\
$R^{TT}$&0.7448 & $4.8\sci{-6}$ & $1.0\sci{-2}$ & \pfactor{1.0} & \pfactor{1.0} \\
$\sigma_{16}^2$&605.9 & $9.2\sci{-5}$ & $2.4\sci{-5}$ & $2.5\sci{-3}$ & \pfactor{1.9} \\
$S_{QO}$ &0.8203& $6.3\sci{-7}$ & $3.8\sci{-6}$ & $1.8\sci{-6}$ & $3.9\sci{-4}$ \\
\hline
\multicolumn{6}{c}{\texttt{SEVEM}}\\
\hline
$S_{1/2}$&1215 & $1.3\sci{-3}$ & \pfactor{0.8} & \pfactor{33} & \pfactor{1.2} \\
$R^{TT}$&0.8194 & $5.6\sci{-5}$ & $5.4 \sci{-2}$ & \pfactor{1.2} & \pfactor{1.0} \\
$\sigma_{16}^2$&583.4 & $6.5\sci{-5}$ & $1.0\sci{-4}$ & $1.6\sci{-3}$ & \pfactor{1.5} \\
$S_{QO}$&0.6547 & $6.3\sci{-5}$ & $2.2\sci{-3}$ & $9.8\sci{-5}$ & $4.1\sci{-2}$ \\
\hline
\multicolumn{6}{c}{\texttt{SMICA}}\\
\hline
$S_{1/2}$ & 1257 & $1.4\sci{-3}$ & \pfactor{0.6}  & \pfactor{25} & \pfactor{1.3} \\
$R^{TT}$ &0.7906 & $2.8\sci{-5}$ & $3.0\sci{-2}$ & \pfactor{1.1} & \pfactor{1.0} \\
$\sigma_{16}^2$&631.0 & $1.4\sci{-4}$ & $1.3\sci{-4}$ & $3.9\sci{-3}$ & \pfactor{1.8} \\
$S_{QO}$&0.8048 & $1.7\sci{-6}$ & $2.9\sci{-5}$ & $6.6\sci{-6}$ & $9.2\sci{-4}$ \\
\end{tabular}
\end{ruledtabular}
\caption{\label{tab:pairwisep}
    $p$-values of the four anomaly statistics ($S_{1/2}$, $R^{TT}$, $\sigma^2_{16}$, $S_{QO}$) for the four \Planck\ component-separated CMB temperature maps (\texttt{Commander}, \texttt{NILC}, \texttt{SEVEM}, and \texttt{SMICA}). 
    For each map: the second column shows the values of the statistics calculated with $S_{1/2}$ given in $\mu\mathrm{K}^4$, $\sigma_{16}^2$ given in $\mu\mathrm{K}^2$, and $R^{TT}$ and $S_{QO}$ both dimensionless;
    the ``diagonal'' entries of the four rows and columns 3--6 are the individual $p$-values of the anomaly statistics in an ensemble of $10^8$ realizations of best-fit \LCDM;
    each off-diagonal entry below the diagonal is the joint $p$-value of the statistic in that row and column;
    the off-diagonal entries above the diagonal (blue) are the factors by which the joint $p$-values exceed the products of the individual $p$-values, i.e., $p(S_A,S_B)/(p(S_A)p(S_B))$.
}
\end{table}

In \cref{tab:S12sig16crossS}, we examine the correlations between the most correlated pair of statistics, $S_{1/2}$ and $ \sigma_{16}^2$, and the next-most correlated statistic $S_{QO}$.  
We find that there is little additional correlation among them, in particular
$0.6\leq p(S_{1/2},\sigma_{16}^2,S_{QO})/(p(S_{1/2},\sigma_{16}^2)p(S_{QO}))\leq 2.1$.

\begin{table}
\begin{ruledtabular}
\begin{tabular}{lcc}
&$S_{1/2}$ and $\sigma_{16}^2$ & $S_{QO}$ \\
\hline
\multicolumn{3}{c}{\texttt{Commander}} \\
\hline
 $S_{1/2}$ and $\sigma_{16}^2$ & 1.2\sci{-4} & \pfactor{1.7} \\
 $S_{QO}$ & 9.1\sci{-7} & $4.4\sci{-3}$ \\
 \hline
\multicolumn{3}{c}{\texttt{NILC}}\\
\hline
 $S_{1/2}$ and $\sigma_{16}^2$ & 9.2\sci{-5} & \pfactor{0.6} \\
 $S_{QO}$ & 2.0\sci{-8} & $3.9\sci{-4}$ \\
 \hline
\multicolumn{3}{c}{\texttt{SEVEM}}\\
\hline
 $S_{1/2}$ and $\sigma_{16}^2$ & 6.5\sci{-5} & \pfactor{1.3} \\
 $S_{QO}$ & 3.6\sci{-6} & $4.1\sci{-2}$ \\
  \hline
\multicolumn{3}{c}{\texttt{SMICA}}\\
\hline
 $S_{1/2}$ and $\sigma_{16}^2$ & 1.4\sci{-4} & \pfactor{2.1} \\
 $S_{QO}$ & 2.7\sci{-7} & $9.2\sci{-4}$ \\
\end{tabular}
\end{ruledtabular}
\caption{\label{tab:S12sig16crossS}For each of the four \Planck\ component-separated CMB temperature maps (\texttt{Commander}, \texttt{NILC}, \texttt{SEVEM}, and \texttt{SMICA}), along the diagonal are 
the joint $p$-value of $S_{1/2}$ and $\sigma_{16}^2$, $p(S_{1/2},\sigma_{16}^2)$, and the $p$-value of $S_{QO}$, $p(S_{QO})$. In the off-diagonal below the diagonal is the joint $p$-value of all three statistics $p(S_{1/2},\sigma_{16}^2,S_{QO})$, while above the diagonal (blue) is the ratio $p(S_{1/2},\sigma_{16}^2,S_{QO})/(p(S_{1/2},\sigma_{16}^2)p(S_{QO}))$. }
\end{table}

Finally, \cref{tab:allfour} gives the joint $p$-values of all four anomaly statistics 
\begin{equation}
    p_{4} \equiv p(S_{1/2},R^{TT},\sigma_{16}^2,S_{QO})
\end{equation}
for the four component-separated maps and the ``correlation factors'' by which they exceed the products of the four individual $p$-values,
\begin{equation}
\frac{p(S_{1/2},R^{TT},\sigma_{16}^2,S_{QO})}
    {p(S_{1/2})p(R^{TT})p(\sigma_{16}^2)p(S_{QO})}\,.
\end{equation}

\begin{table}
\begin{ruledtabular}
\begin{tabular}{lcc}
Map &  $p_4$ & Correlation Factor \\
\hline
\texttt{Commander}&  $3 \times 10^{-8}$ & 51
    \\
\texttt{NILC}&  $<1 \times 10^{-8}$  & N/A
    \\
\texttt{SEVEM}& $18 \times 10^{-8}$ & 40
    \\
\texttt{SMICA}& $1 \times 10^{-8}$ & 64
\end{tabular}
\end{ruledtabular}
\caption{\label{tab:allfour}The joint $p$-values, $p_4$, of $S_{1/2}$, $R^{TT}$, $\sigma_{16}^2$, and $S_{QO}$, and the correlation factors by which the joint $p$-values exceed the products of the four individual $p$-values, estimated on ensembles of $10^8$ realizations of \Planck\ best-fit \LCDM\ for the four \Planck\ component-separated CMB temperature maps (\texttt{Commander}, \texttt{NILC}, \texttt{SEVEM}, and \texttt{SMICA}).}
\end{table}

We see in \cref{tab:allfour} that $p_4$ for \texttt{Commander}, \texttt{NILC}, and \texttt{SMICA} are all $\leq 3\times 10^{-8}$, corresponding to  $>5.4\sigma$ significance! Even the one outlier, \texttt{SEVEM}, has a $p_4$ corresponding to  $\sim5.1\sigma$. There is no particular reason to trust the \texttt{SEVEM} result over the others, and if it were the correct value the other three would be extremely unlikely ($<1/10^{18}$) to have happened by chance -- there would have to be some reason why all three of them come out so much smaller. We reiterate that this $3\times 10^{-8}$ result is not from multiplying each statistic's individual $p$-value. This is the actual rate at which \LCDM\ reproduces the same anomalies in 100,000,000 CMB realizations.

The correlation factor -- the factor by which $p_4$ exceeds the product of the four individual $p$-values -- is thus consistently less than $100$, so that $p_4$ is vastly smaller than the minimum of the individual $p$-values.
Most of that correlation factor is accounted for by the correlation between $S_{1/2}$ and $\sigma^2_{16}$, 
but recall that even $p(S_{1/2},\sigma^2_{16}) < \frac{1}{10} \mathrm{min}(p(S_{1/2}),p(\sigma^2_{16}))$.

As remarked above, here we have worked with CMB-only simulated sky maps, i.e., we did not add simulated noise to our realizations of the CMB sky. This is motivated by the fact that the \Planck\ (and \WMAP) measurements of large-scale CMB temperature fluctuations are almost completely cosmic-variance-limited so that we do not expect the noise to significantly affect our results and conclusions. To confirm this, we conducted the same study on the \Planck\ FFP10 simulations \cite{Planck:2015txa} and found nearly identical results. This is further supported by the fact that our calculated $p$-values of anomaly statistics agree with the corresponding \Planck\ published values. Additionally, WMAP and \Planck, with very different noise structures and properties, have both detected the large-scale anomalies with very similar individual $p$-values.  Given these points, it is very hard to see how the noise would be responsible for all the anomalies together. Working with noise-free simulations makes it possible to create ensembles of $10^8$ realizations, which would not have been possible if we had added noise to the simulations.

We have also not attempted to simulate residual foregrounds.  
The \texttt{Commander}, \texttt{NILC}, \texttt{SEVEM}, and \texttt{SMICA} teams have done that better than we can hope to. It seems extraordinarily unlikely that all four of these foreground removal methodologies would leave behind the identical anomalies, with nearly the same values on all of the \Planck\ maps (and similarly for all the foreground-removed \WMAP\ maps).

\para{Conclusions.}
The joint significance of the four statistics we chose to consider much exceeds $5\sigma$, and for three of four maps exceeds $5.5\sigma$. There are many caveats when interpreting this result.  Of course, one would expect a result this significant if we were to simply multiply each statistic's individual $p$-value, but this is explicitly not what we have done. If the statistics were correlated, $p_4$ may have just as well been the minimum of the 4 individual $p$-values. The fact that $p_4$ is indeed much lower informs our conclusion: under standard \LCDM, the large-angle anomalies are almost entirely uncorrelated. If there is a single physical effect or a systematic that causes these features, it is not currently known or captured within the standard model. 

If we are to make claims about the implication of this resulting $p_4$, we should certainly mitigate it by the effect of some look-elsewhere penalty,  although, it is not obvious how one would quantify this. 
However, the typical justification of a $5\sigma$ threshold for discovery is that it overwhelms even strong look-elsewhere concerns. We believe that our findings, even while considering look-elsewhere effects, are strong evidence of the need for further study into the source of the anomalies.

One could conduct a more comprehensive study of the many statistics that point toward the violation of statistical isotropy.
The {\it Planck} isotropy and statistics papers \cite{Planck:2013lks,Planck:2015igc,Planck:2019evm} contain many statistics to test statistical isotropy, many of them presumably not independent of one another.
Of these {\it Planck} statistics, the anomalous ones were a  substantial fraction of the statistics that they explored.
The four statistics we have included here are, deliberately, the same ones chosen by \rcite{Muir:2018hjv}.  
Like us, they were aiming for a representative subset of four commonly discussed classes of anomaly that \textit{a priori} seem to represent four physically different ways in which the CMB sky appeared to be anomalous, and specifically to demand the violation of statistical isotropy (despite the caveats discussed above).
These four were also representatives of the earliest and simplest tests of isotropy applied to the WMAP and {\it Planck} data sets.

We have omitted many other statistics that are likely to be tightly correlated with these four in $\Lambda$CDM.
But we have also omitted statistics that are both anomalous and unlikely to be correlated with these.
Consider, e.g., the observed $S_{1/2}$ anomaly for $E$-mode polarization \cite{Chiocchetta:2020ylv}, the anomalously low large-angle skewness and high large-angle kurtosis \cite{Planck:2013lks}, the anomalous alignment between multipole vectors of $\ell=3$ and $\ell=5$ \cite{Land:2005ad} and even up to $\ell=9$  \cite{Copi_2004}. 
We expect that in \LCDM\ these are all nearly statistically independent of the four statistics we have considered and would therefore increase the joint significance if added to our analysis.

Look-elsewhere penalties should no doubt be paid for statistics that have been considered and are not anomalous (but are independent of one another), but only while properly accounting for all the ones that are anomalous (and independent of one another).
There is no look-elsewhere penalty for failing to include representatives of classes of anomalous statistics that are not expected to be  tightly correlated with our statistics. Moreover, look-elsewhere penalties cannot possibly be paid for statistics that were never considered; nor is it appropriate to devise a catalog of new statistics that might have been considered in order to dilute the significance of existing statistics with ever-larger look-elsewhere penalties.

Put another way, there are both look-elsewhere penalties due to non-anomalous statistics and ``look-more-closely rewards'' due to anomalous statistics that are uncorrelated with the ones we have included. 
The latter seem to outweigh the former, with each additional uncorrelated anomaly requiring look-elsewhere penalties from many additional non-anomalous statistics to offset the decrease in the joint $p$-value it would imply.
This is just what one would expect if a physical effect is the cause of these anomalies.

Finally, regarding look-elsewhere penalties, one should understand 
the reason for adopting $5\sigma$ as a standard.  It is not because we really believe we need to reduce the probability of a fluke to parts in $10^7$ before accepting a result.
As Lyons writes in “Five Sigma Revisited” \cite{Lyons2023FiveSigma},
\begin{quote}
    There are several reasons why the stringent $5\sigma$ rule is used (in particle physics). The first is that it provides some degree of protection against falsely claiming the observation of a discrepancy with the Standard Model. 
    There have been numerous $3\sigma$ and $4\sigma$ effects in the past that have gone away when more data was collected.
\end{quote}
We conclude that it is very unlikely that we live in a realization of a Gaussian random statistically isotropic \LCDM\ universe,
and this is probably not because of the violation of Gaussianity.
The anomalies should not be dismissed and may very well be shedding light into a direction that alternative models can aim to predict.

\begin{acknowledgments}
We thank the referees for useful comments that motivated us in particular to increase the discussion of look-elsewhere penalties, and to modify the title.
This work made use of the High Performance Computing Resource in the Core Facility for Advanced Research Computing at Case Western Reserve University.
Some of the results in this paper have been derived using the \texttt{healpy} and \texttt{HEALPix} packages  \cite{Gorski:2004by,Zonca2019}. J.J. acknowledges partial support by the Jump Trading Graduate Fellowship. J.J., C.J.C., and G.D.S.\ acknowledge partial support from NASA ATP grant RES240737; G.D.S.\ from DOE grant DESC0009946.
Y.A.\ acknowledges support by the Spanish Research Agency (Agencia Estatal de Investigaci\'on)'s grant RYC2020-030193-I/AEI/10.13039/501100011033, by the European Social Fund (Fondo Social Europeo) through the  Ram\'{o}n y Cajal program within the State Plan for Scientific and Technical Research and Innovation (Plan Estatal de Investigaci\'on Cient\'ifica y T\'ecnica y de Innovaci\'on) 2017-2020, and by the Spanish Research Agency through the grant IFT Centro de Excelencia Severo Ochoa No CEX2020-001007-S funded by MCIN/AEI/10.13039/501100011033.
\end{acknowledgments}

\bibliography{main}

\end{document}